\documentclass{jfm}
\usepackage{graphicx}
\usepackage{epstopdf, epsfig}
\usepackage{bm}
\usepackage{braket}
\usepackage{array}

\shorttitle{Simulation of internal wave-driven turbulent mixing}
\shortauthor{Y. Onuki, S. Joubaud and T. Dauxois}

\title{Simulating turbulent mixing caused by local instability of internal gravity waves}

\author{Yohei Onuki\aff{1}
  \corresp{\email{onuki@riam.kyushu-u.ac.jp}},
  Sylvain Joubaud\aff{2}$^,$\aff{3}
 \and Thierry Dauxois\aff{2}}

\affiliation{\aff{1}Research Institute for Applied Mechanics, Kyushu University,
6-1 Kasuga-koen, Kasuga, Fukuoka, Japan
\aff{2}Universit\'e de Lyon, ENS de Lyon, Universit\'e Claude Bernard, CNRS,\\ Laboratoire de Physique, F-69342 Lyon, France
\aff{3}Institut Universitaire de France, France}

\begin{document}

\maketitle

\begin{abstract}
With the aim of assessing internal wave-driven mixing in the ocean, we develop a new technique for direct numerical simulations of stratified turbulence.  Since the spatial scale of oceanic internal gravity waves is typically much larger than that of turbulence, fully incorporating both in a model would require a high computational cost, and is therefore out of our scope.  Alternatively, we cut out a small domain periodically distorted by an unresolved large-scale internal wave and locally simulate the energy cascade to the smallest scales.  In this model, even though the Froude number of the outer wave, $Fr$, is small such that density overturn or shear instability does not occur, a striped pattern of disturbance is exponentially amplified through a parametric subharmonic instability.  When the disturbance amplitude grows sufficiently large, secondary instabilities arise and produce much smaller-scale fluctuations.  Passing through these two stages, wave energy is transferred into turbulence energy and will be eventually dissipated.  Different from the conventional scenarios of vertical shear-induced instabilities, a large part of turbulent potential energy is supplied from the outer wave and directly used for mixing.  The mixing coefficient $\Gamma=\epsilon_P/\epsilon$, where $\epsilon$ is the dissipation rate of kinetic energy and $\epsilon_P$ is that of available potential energy, is always greater than 0.5 and tends to increase with $Fr$.  Although our results are mostly consistent with the recently proposed scaling relationship between $\Gamma$ and the turbulent Froude number, $Fr_t$, the values of $\Gamma$ obtained here are larger by a factor of about two than previously reported.
\end{abstract}

\section{Introduction}
Diapycnal mixing in the mid-depth and deep ocean is a major contributor to the global overturning circulation that transports water mass, heat, and various biochemical substances \citep{munk_abyssal_1998}.  Therefore, quantifying and parameterizing the small-scale turbulent mixing that cannot be resolved in ocean circulation models are crucial to our ability of predicting changes in the Earth's environment \citep{mackinnon_climate_2017}.

The difficulty in understanding interior ocean mixing comes from the multi-scale nature of the system.   The smallest length scale of the fluid motion that enhances heat conduction or salinity diffusion is normally $O(1 {\rm cm})$ or much smaller \citep{thorpe_turbulent_2005}.  Energy of this turbulent motion is mainly supplied from internal gravity waves, whose spatial scale is the order of tens or hundreds of meters.  To fully assess the mixing in the ocean using a three-dimensional direct numerical simulation (DNS) model, accordingly, would need to incorporate larger than $O(10^{12})$ grid points, which is an extremely challenging task even with the latest parallel computing systems.

Since the ordinary DNS cannot resolve both the typical scales of waves and turbulence, many studies simplify the problem.  A widely considered situation is the mixing occurring in a vertically sheared horizontal flow due to Kelvin-Helmholtz or Holmboe instabilities \citep[see, e.g.,][]{smyth_efficiency_2001, salehipour_turbulent_2016, dauxois_confronting_2020}.  In this case, the energy of turbulence is supplied from the kinetic energy of a background horizontal flow and will be redistributed to viscous dissipation and vertical buoyancy flux.  In a stationary state, the vertical buoyancy flux coincides with the conversion rate from the available potential energy to the background potential energy.  Since \cite{osborn_estimates_1980}, this energy balance has been regarded as representative in the ocean.  Great attention has been directed to the ratio between the turbulence energy loss and the background potential energy gain, or the so-called mixing efficiency \citep{gregg_mixing_2018}, which is a fundamental parameter that influences the global-scale ocean circulation.

It is true that the ocean interior is typically dominated by slowly varying vertically sheared horizontal flows accompanying geostrophic currents or near-inertial waves, to which the above-mentioned model may well apply.  However, this simplified model does not necesarily portrays high-frequency wave-driven mixing.  In general, internal wave energy is divided into the kinetic energy and the available potential energy.  If rotation is excluded, these two are evenly partitioned.  As was shown in laboratory experiments of Rayleigh-Taylor instability, when mixing is caused by the release of available potential energy through convection, the mixing efficiency becomes several factors higher than that in the cases of shear instability \citep{davies_wykes_efficient_2014}.  Accordingly, to fully understand the mixing caused by internal waves and to properly parameterise the mixing efficiency, it is vital to extend the discussion from classical shear instability to a wider variety of scenarios for turbulence generation.

Most of the previous numerical studies that examined turbulence generated by internal waves \citep[such as,][]{lombard_breakdown_1996, bouruet-aubertot_particle_2001, fritts_mean_2006, achatz_gravity-wave_2007, fritts_gravity_2009-3, fritts_gravity_2009-1, fritts_gravity_2009-2, fritts_numerical_2016} have resolved both the scales of waves and turbulence in a model.  In this study, on the other hand, with the aim of increasing the Reynolds number while reducing the computational cost, we do not resolve the largest wave component.  Instead, we cut out a small domain periodically distorted by an unresolved outer wave, and simulate the excitation and dissipation of turbulence within it.  It is noted that this configuration may seem similar to \cite{inoue_efficiency_2009}.  They tilted the angle of a rigid model domain to imitate forcing from a larger-scale internal wave.  Here, in contrast, we distort the shape of the domain to take into account the effects of oscillating velocity shear and buoyancy gradient that play essential roles on the energy conversion from waves to turbulence.

In the present model, the wave field is idealised as a linear shear flow, such that the turbulence enhanced in it is statistically homogeneous.  Hence, we can assume periodic boundary conditions in all directions and employ the Fourier spectral discretisation, enabling precise representation of an energy cascade to the smallest dissipation scale.  As for homogeneous turbulence of stratified shear flow, there have been numerous studies using DNS \citep[][and references therein]{portwood_study_2019}.  They utilised a generic technique originally invented by \cite{rogallo_numerical_1981} that we basically follow.  However, the present study differs from the existing ones in two points:

1. In our model, the background shear varies periodically associated with the oscillation of the outer wave.  As we shall see, this periodic variation causes parametric excitation of disturbances.  Consequently, the turbulence energy is more easily amplified than in the conventional cases of stationary shear flows.

2. Since the buoyancy gradient is also changed by the outer wave, a fluid parcel crossing this gradient at some instant gains the available potential energy, which subsequently drives turbulent motion.  This contrasts with the classical view of shear-driven mixing in a horizontal flow that is associated only with the shear production of kinetic energy.

In general, properties of stratified turbulence, including the mixing efficiency, are determined by the competing effects of buoyancy, inertia, viscosity and diffusion.  Therefore, a thorough analysis requires to check the dependency of the experimental results on at least three dimensionless parameters.  Among these, motivated by the suggestion of \cite{maffioli_mixing_2016} that competition between inertia and buoyancy is primarily important in high-Reynolds number regimes, we vary the Froude number of the outer wave that controls the energy level of turbulence while fixing the other parameters.

The paper is organised as follows.  The model and the calculation methods are described in \S2.  We present the simulation results in \S3 and discuss the mixing efficiency from an energetic viewpoint in \S4.  Concluding remarks are presented in \S5.

\section{Formulation and calculation method}
In this study, we consider a stably stratified fluid motion governed by the Navier-Stokes equation with the Boussinesq approximation, specifically written as
\begin{subequations} \label{eq:governing}
\begin{eqnarray}
\frac{\partial \bm{u}}{\partial t} + \bm{u} \cdot \nabla \bm{u} & = & - \nabla p + N b \ \bm{e}_v + \nu \nabla^2 \bm{u}  \\
\frac{\partial b}{\partial t} + \bm{u} \cdot \nabla b & = & - N \bm{u} \cdot \bm{e}_v + \kappa \nabla^2 b ,
\end{eqnarray}
\end{subequations}
where $\bm{u}$ is the three-dimensional velocity vector satisfying the incompressible condition, $\nabla \cdot \bm{u} = 0$, $p$ is the pressure divided by the reference density, $b$ is the buoyancy, $N$ is the background buoyancy frequency set as a constant, $\nu$ is the kinematic viscosity, $\kappa$ is the diffusivity, and $\bm{e}_v$ is the unit vector pointing upwards.  Here, $b$ is scaled so as to have the unit of velocity and related to the fluid density, $\rho$, as $\rho = \rho_{ref}(1 - N^2 X_3/g - N b/g)$, where $g$ is the acceleration of gravity, $\rho_{ref}$ is the reference density, and $X_3$ is the vertical coordinate.

\begin{figure}
  \centerline{\includegraphics[bb=0 0 1226 407, width=\textwidth]{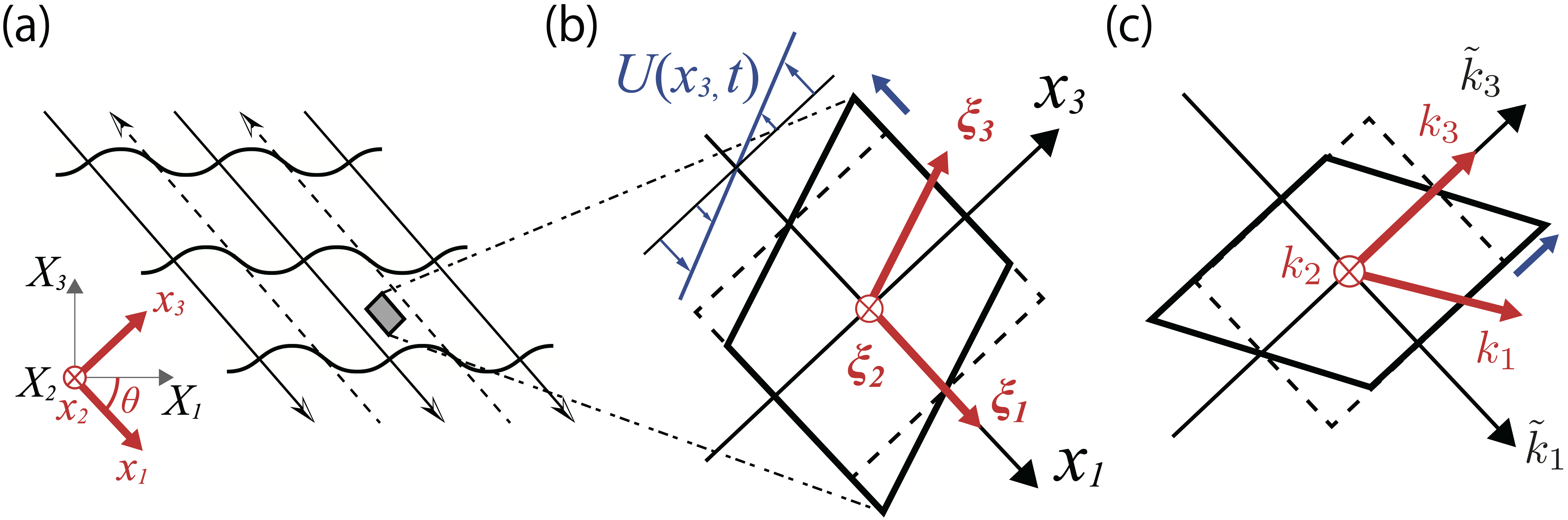}}
\caption{(a) We consider an internal gravity wave obliquely propagating in a stratified fluid.  Thick black curves represent the isopycnal surfaces and the arrows indicate the direction of the flow velocity.  A cartesian coordinate system, $(x_1, x_2, x_3)$, is arranged such that $x_1$ points to the direction of the flow velocity, $x_3$ to the velocity gradient, and $x_2$ perpendicular to them.   (b) We cut out a small domain from the internal wave field, in which the velocity $U$ is linear in $x_3$ and oscillates in time.  A time-dependent coordinate system, $(\xi_1, \xi_2, \xi_3)$, that follows the background flow velocity is introduced.  The shape of the model domain (thick parallelogram) is fixed in the $\xi$ frame and hence periodically distorted.  (c) The wavenumber coordinates, $(k_1, k_2, k_3)$, which are defined by taking Fourier transform with respect to $(\xi_1, \xi_2, \xi_3)$, also vary with time.}
  \label{fig:coordinate}
\end{figure}

We envisage a plane internal gravity wave propagating in a direction with an angle~$\theta$ against the vertical direction.  Let us assign the $x_3$-axis to this direction, the $x_1$-axis to the direction of the flow velocity, and the $x_2$-axis perpendicular to them (figure \ref{fig:coordinate}a).  In the following, velocity components along the $x_1$-, $x_2$-, and $x_3$-axes are respectively represented as $u_1$, $u_2$, and $u_3$.  We may write the ordinary coordinates $(X_1, X_2, X_3)$, with $X_3$ pointing to $\bm{e}_v$, as $X_1 = x_1 \cos\theta + x_3 \sin\theta, \ X_2 = x_2, \ X_3 = - x_1 \sin\theta + x_3 \cos\theta$.  Now, we assume $\nu = \kappa$ for simplicity, which allows a plane wave solution of (\ref{eq:governing}) to be written down as
\begin{subequations}
\begin{eqnarray}
u_1 & = & U_0 e^{- \nu k^2 t} \sin(kx_3 - \omega t + \alpha) \equiv U(x_3, t)  \label{eq:wave-U}  \\
b & = & U_0 e^{- \kappa k^2 t} \cos(kx_3 - \omega t + \alpha) \equiv B(x_3, t)  \label{eq:wave-B}  \\
p & = & \frac{N U_0 \cos\theta}{k} e^{- \nu k^2 t} \sin(kx_3 - \omega t + \alpha) \equiv P(x_3, t) \\
\omega & = & N \sin\theta  \label{eq:dispersion}
\end{eqnarray}
\end{subequations}
and $u_2 = u_3 = 0$, where $U_0$ is the amplitude of the wave velocity, $k$ is the wavenumber, and $\alpha$ is the initial phase factor that is arbitrarily chosen.

Next, we cut out a small domain from this wave field.  By assuming that the wavelength is sufficiently larger than the domain size, i.e., taking the asymptotic limit of small $k$, we expand (\ref{eq:wave-U}) and (\ref{eq:wave-B}) as
\begin{subequations}\label{eq:wave}
\begin{eqnarray}
U & = & - U_0 \sin \left( \omega t - \alpha \right) + \underbrace{S_0 \cos \left( \omega t - \alpha \right)}_{\equiv S(t)} x_3 + O(k^2)  \\
B & = & U_0 \cos \left( \omega t - \alpha \right) + \underbrace{S_0 \sin \left( \omega t - \alpha \right)}_{\equiv M(t)} x_3 + O(k^2) ,
\end{eqnarray}
\end{subequations}
where $S_0 \equiv U_0 k$ represents the maximum velocity shear, and $S$ and $M$ are instantaneous velocity shear and buoyancy gradient.  In this way, the plane wave solution can be locally reduced to an oscillating linear shear flow (figure \ref{fig:coordinate}b).  The frequency of oscillation, $\omega$, is linked to the propagation angle $\theta$ via the dispersion relation of internal gravity waves (\ref{eq:dispersion}).

Finally, we consider disturbances superimposed on this wave solution.  Variables are rewritten as $(u_1, u_2, u_3, b, p) = \left(U + u_1', u_2', u_3', B + b', P + p' \right)$.  Here, the disturbances are advected by the space- and time-dependent background flow, $U(x_3, t)$.  To eliminate this passive advection effect from consideration, we introduce a new coordinate system that moves following the background velocity as $\xi_1 = x_1 - \int U dt, \xi_2 = x_2, \xi_3 = x_3$ (see, again, figure \ref{fig:coordinate}b).  The governing equations for the disturbance components are, consequently,
\begin{subequations} \label{eq:governing2}
\begin{eqnarray}
\frac{\partial u'_i}{\partial t} + \frac{\partial u'_j u'_i}{\partial \tilde{\xi}_j}  + S \delta_{i1} u'_3 & = & - \frac{\partial p'}{\partial \tilde{\xi}_i} + N b' (\cos\theta \delta_{i3} - \sin\theta \delta_{i1}) + \nu \frac{\partial^2 u'_i}{\partial \tilde{\xi}_j \partial \tilde{\xi}_j}  \\
\frac{\partial b'}{\partial t} + \frac{\partial u'_j b'}{\partial \tilde{\xi}_j} + M u'_3 & = & - N (\cos\theta u'_3 - \sin\theta u'_1) + \kappa \frac{\partial^2 b'}{\partial \tilde{\xi}_j \partial \tilde{\xi}_j}  \\
\frac{\partial u'_j}{\partial \tilde{\xi}_j} & = & 0 ,
\end{eqnarray}
\end{subequations}
where $\partial / \partial \tilde{\xi}_i \equiv \partial / \partial \xi_i - \int S dt \delta_{i3} \partial / \partial \xi_1$, $i$ and $j$ represent $1, 2$ or $3$, and the summation convention for repeated indices is used.

It is noteworthy that, when we neglect the nonlinear, viscosity, and diffusion terms, the system of equations (\ref{eq:governing2}) reduces to the model proposed by \cite{ghaemsaidi_three-dimensional_2019}.  As shown by their stability analysis, of this model, several kinds of instability occur depending on the angle and the amplitude of the outer wave.  Thus, the energy of the system is spontaneously amplified even without any external forcing term.  Here, we shall call this process ``local instability'' of internal waves.

In the present system, budgets of the kinetic energy and the available potential energy are governed by
\begin{subequations}
\begin{eqnarray}
\frac{d}{dt} \underbrace{\Braket{ \frac{u'_i u'_i}{2} }}_{\mathcal{E}_K} & = & \underbrace{- S \Braket{u'_1 u'_3}}_{\mathcal{P}_K} \underbrace{+ N \cos\theta \Braket{u'_3 b'} - N \sin\theta \Braket{u'_1 b'}}_{\mathcal{C}} - \underbrace{\nu \Braket{ \frac{\partial u'_i}{ \partial \tilde{\xi}_j} \frac{\partial u'_i}{ \partial \tilde{\xi}_j} }}_{\epsilon}   \\
\frac{d}{dt} \underbrace{\Braket{ \frac{b'^2}{2} }}_{\mathcal{E}_P} & = & \underbrace{- M \Braket{u'_3 b'}}_{\mathcal{P}_P} \underbrace{- N \cos\theta \Braket{u'_3 b'} + N \sin\theta \Braket{u'_1 b'}}_{- \mathcal{C}} - \underbrace{\kappa \Braket{ \frac{\partial b'}{\partial \tilde{\xi}_j} \frac{\partial b'}{\partial \tilde{\xi}_j} }}_{\epsilon_P} ,
\end{eqnarray}
\end{subequations}
where $\braket{ \ }$ denotes the spatial averaging over the whole domain.  From these expressions, we understand that the kinetic energy, $\mathcal{E}_K$, and the available potential energy, $\mathcal{E}_P$, are produced by the terms $\mathcal{P}_K$ and $\mathcal{P}_P$, respectively, exchanged through the term $\mathcal{C}$, and eventually dissipated in heat and background potential energies, at the rate of $\epsilon$ and $\epsilon_P$, respectively.  At variance with the vertically sheared horizontal flow cases, the sign of the $\mathcal{E}_K$-$\mathcal{E}_P$ conversion rate $\mathcal{C}$ is not determined a priori; it depends on the relative magnitudes of production and dissipation terms.

In this study, equations (\ref{eq:governing2}) are numerically solved in the wavenumber domain $\bm{k}$ using the Fourier spectral method by assuming triply periodic boundary conditions with respect to $\xi_j$ (not to $X_j $ or $x_j$).  Since the physical coordinates $\xi_j$ now explicitly depend on time, the wavenumber coordinates also vary with time; specifically, when we write the wavenumber in the fixed frame as $\tilde{\bm{k}} = (\tilde{k}_1, \tilde{k}_2, \tilde{k}_3)$, the wavenumber for the moving frame $\bm{k} = (k_1, k_2, k_3)$ follows $k_1=\tilde{k}_1, k_2=\tilde{k}_2, k_3 = \tilde{k}_3 + \int S dt \tilde{k}_1$ (figure \ref{fig:coordinate}c).  For the calculation, we basically follow the procedure by \cite{chung_direct_2012}; the viscous and diffusive terms are analytically solved by the integration factor method, the pressure terms are diagnosed at each step enforcing the incompressibility constraint.  The remaining terms are integrated using the low-storage third-order Runge-Kutta scheme of \cite{spalart_spectral_1991}.  The aliasing errors are eliminated by the combination of wavenumber truncation and phase shifting.  Different from the stationary shear cases, we do not remesh the grid during the calculation because the effect from grid skewing would be minor compared to the aliasing error arising from remeshing in the present settings.  The calculation domain is a cubic box and the grid spacing is equivalent in all directions.  To ensure the stability of calculation, as well as to sufficiently resolve buoyancy oscillation, the time step $\Delta t$ is controlled via a modified form of the Courant-Friedrichs-Lewy condition such that $\Delta t  \leq {\rm min} (1.83 / {\rm max}(\bm{u} \cdot \bm{k}), 0.1 / N)$.

\begin{figure}
  \centerline{\includegraphics[bb=0 0 503 517, width=0.4\textwidth]{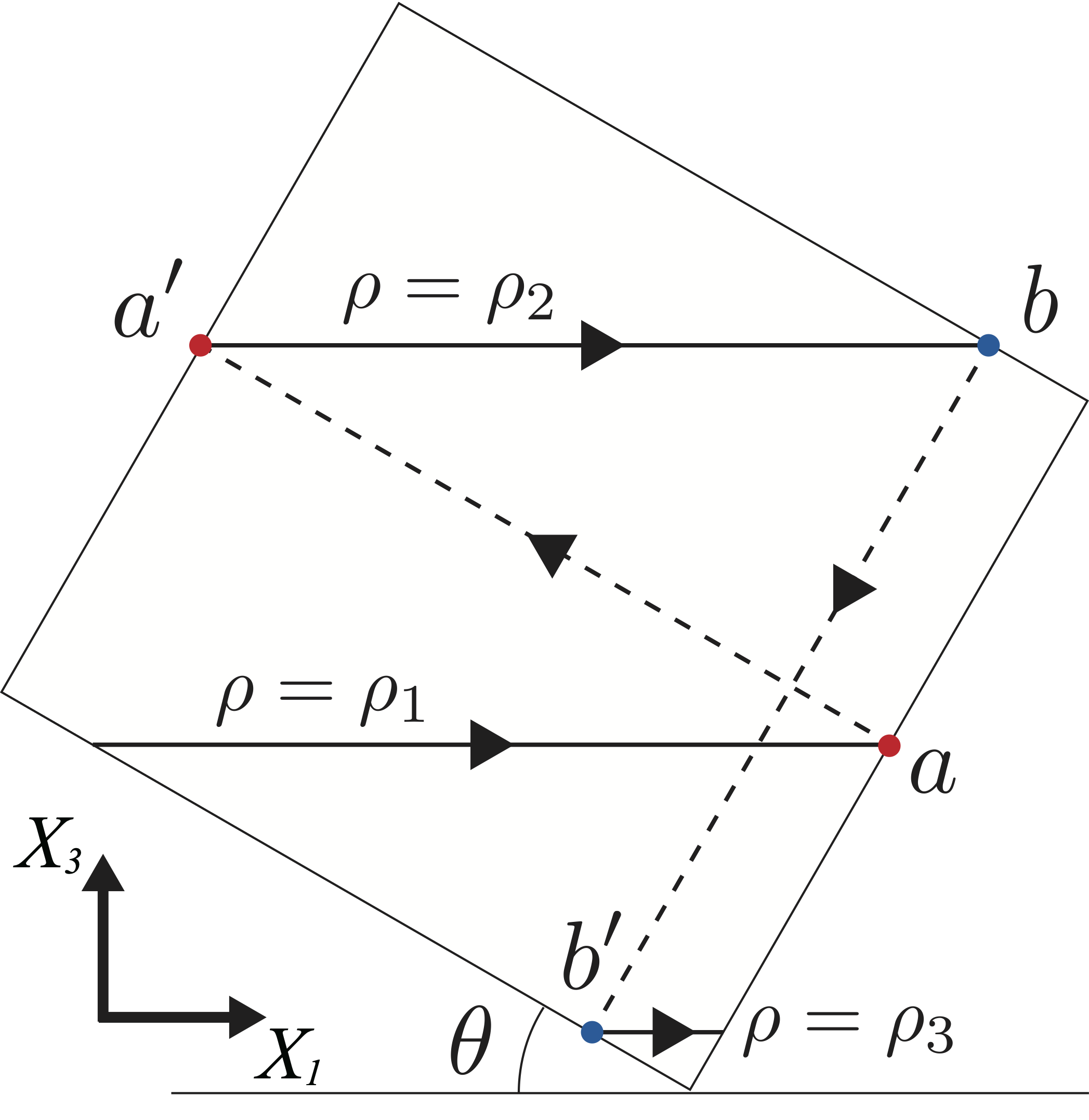}}
\caption{A side view of the model domain.  The periodic boundary condition connects points, $a$ to $a'$ and $b$ to $b'$.  Accordingly, the isopycnal surfaces $\rho = \rho_1$, $\rho = \rho_2$ and $\rho=\rho_3$ are connected with each other.}
  \label{fig:square}
\end{figure}

Generally speaking, in stratified flows, vertically sheared large-scale horizontal motions emerge from a turbulent state \citep[e.g.,][]{chung_direct_2012}.  However, in our model, growth of such a flow structure is suppressed due to the following reason.  As shown in figure \ref{fig:square}, since the model domain is tilted and the periodic boundary condition connects each face to its opposite side, an isopycnal surface at some level connects to that at another level, such that $a \to a'$ and $b \to b'$.  When we extend the area of this surface, it will fill in the whole domain if $\tan \theta$ is an irrational number.  In this way, all the levels are kinematically connected.  Therefore, this model does not allow the existence of a vertically sheared horizontally homogeneous motion.  This thought also applies to the density distribution.  As is known, a turbulent patch in a homogeneously stratified fluid locally mixes the density gradient to generate a staircase structure.  Then, signals of the disrupted stratification propagate along an isopycnal surface.  In the present model, since all the isopycnal surfaces are connected through the periodic boundaries, the staircase structure is dispersed with time and eventually homogenised.  From this property, we can conduct experiments without being concerned with the changes in the background stratification.

The non-dimensional control parameters in the experiments are the external Froude number, $Fr = S_0 / N$, the external Reynolds number, $Re = S_0L^2 / \nu$, the Prandtl number, $Pr = \nu / \kappa$, and the wave frequency divided by the buoyancy frequency, $\omega / N$.  Here, we have introduced the domain length $L$, which is fixed to be $2 \pi$ in our model.  For thermally stratified ocean, $Pr$ is around 7 but for simplicity it is set to unity here.  The Reynolds number and the Froude number are now defined based on the external conditions and should be distinguished from those determined by the length and velocity scales of turbulence.  In this study, we fix $Re$ to be 300000 but vary $Fr$ from 0.1 to 1.2.  The wave frequency is set relatively high, $\omega / N = 0.6$, to avoid excessive distortion of the domain shape.  The initial condition is a small isotropic Gaussian white noise added both for the velocity and buoyancy fields.  According to \cite{ghaemsaidi_three-dimensional_2019}, although the density overturn or the shear instability does not occur in the present parameter regime, disturbances are expected to be amplified through parametric subharmonic instability (PSI).  To reduce the computation cost, the total number of grid points is first set as $512^3$ with the spherical truncation of wavenumbers at the $241$ mode, and later raised up to $1024^3$ with the truncation at the $482$ mode before the disturbance energy is fully developed.

A basic theory of PSI tells us that the instability growth rate is proportional to the amplitude of the background wave \citep{sonmor_toward_1997}, which corresponds to $Fr$ in the present case.  Therefore, the time required for the energy of the system to grow up would be roughly proportional to $Fr^{-1}$.  We accordingly vary the total calculation time, $t_{\rm end}$, as listed in Table 1.  To quantify statistical properties of turbulence, we average data over $t_{\rm end} - T < t \leq t_{\rm end}$, where $T \equiv 2\upi / \omega$ is the period of the outer wave.  It is noted that, although the system is expected to eventually reach a quasi-stationary state in each run, the high computational cost inhibits us from conducting experiments over a so long period.  We inevitably admit the existence of some deviations in our data from the genuine long-term means.

To ensure the sufficiency of model resolution, we check the Kolmogorov scale $\eta_K = (\nu^3 / \epsilon)^{1/4}$ and the Ozmidov scale $\eta_O = (\epsilon / N^3)^{1/2}$, which are the lower and upper bounds of the range of isotropic turbulence.  Then, we confirm that both are resolved in all experiments.  For example, we show in Table 1 the values of $\eta_K k_{max}$, the Kolmogorov scale multiplied by the maximum wavenumber, which should be larger than 1 according to \cite{de_bruyn_kops_direct_1998}.

\begin{table}
  \begin{center}
\def~{\hphantom{0}}
  \begin{tabular}{m{3em}m{3em}m{3em}m{3em}ccc}
      $Fr$ & $t_{\rm end}$ & $Re_b$ & $Fr_t$ & $ \epsilon_P / (\epsilon + \epsilon_P)$ & $\mathcal{P}_P / (\mathcal{P}_K + \mathcal{P}_P)$ & $\eta_K k_{max}$ \\[3pt]
       0.1 & $85T$ & 1.55 & 0.014 & 0.41 & 0.54 & 1.6  \\
       0.2 & $40T$ & 4.74 & 0.036 & 0.38 & 0.55 & 1.7 \\
       0.3 & $32T$ & 62.6 & 0.066 & 0.35 & 0.53 & 1.1 \\
       0.4 & $25T$ & 32.5 & 0.078 & 0.34 & 0.61 & 1.5 \\
       0.5 & $19T$ & 86.9 & 0.099 & 0.36 & 0.56 & 1.3 \\
       0.6 & $16T$ & 47.8 & 0.105 & 0.37 & 0.62 & 1.6 \\
       0.7 & $13T$ & 43.1 & 0.057 & 0.37 & 0.78 & 1.8 \\
       0.8 & $12T$ & 188 & 0.088 & 0.40 & 0.95 & 1.3 \\
       0.9 & $11T$ & 475 & 0.178 & 0.45 & 0.70 & 1.1 \\
       1.0 & $9T$ & 611 & 0.212 & 0.44 & 0.67 & 1.1 \\
       1.1 & $8T$ & 373 & 0.090 & 0.47 & 1.08 & 1.3 \\
       1.2 & $7T$ & 571 & 0.173 & 0.47 & 0.61 & 1.2 \\
  \end{tabular}
  \caption{Values of the control parameter $Fr \equiv S_0/N$, the total calculation time $t_{\rm end}$, and the dimensionless quantities obtained in each run.  Here, $Re_b \equiv \epsilon / (\nu N^2)$ is the buoyancy Reynolds number, $Fr_t \equiv \epsilon / (N \mathcal{E}_K)$ is the turbulent Froude number, and $\eta_K \equiv (\nu^3 / \epsilon)^{1/4}$ is the Kolmogorov scale.  To obtain $\epsilon$, $\epsilon_P$, $\mathcal{E}_K$, $\mathcal{P}_K$, and $\mathcal{P}_P$, temporal averages are taken over $t_{\rm end} - T < t \leq t_{\rm end}$.}
  \label{tab:kd}
  \end{center}
\end{table}

\section{Results}
\begin{figure}
  \centerline{\includegraphics[bb=0 0 701 628, width=\textwidth]{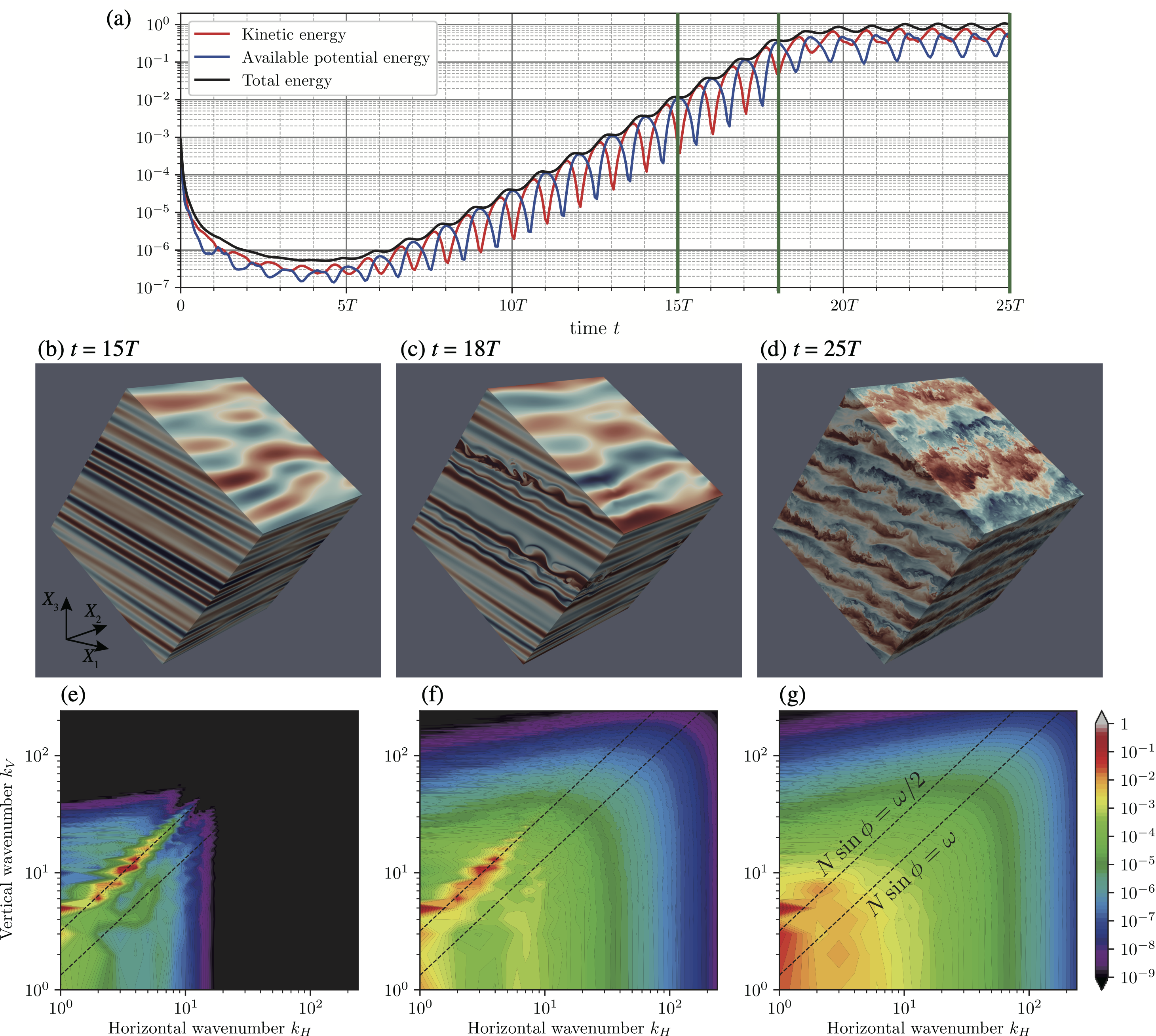}}
\caption{Results of the experiment with $Fr=0.4$. (a) Time series of the kinetic energy (red), the available potential energy (blue), and their sum (black).  They are normalised such that the total energy becomes 1 at the end of the experiment.  The vertical green lines indicate $t = 15T, 18T$, and $25T$, the times that the following panels correspond to.  (b)-(d) Buoyancy perturbation $b'$ on the surface of the calculation domain at $t=15T$(b), $18T$(c), and $25T$(d).  Red and blue correspond to positive and negative values.  (e)-(g) Energy spectra in the horizontal and vertical wavenumber space, $E(k_H, k_V)$, at $t=15T$(e), $18T$(f), and $25T$(g).  Here, $k_H$ indicates the wavenumber against the horizontal directions, $X_1$ and $X_2$, and $k_V$ indicates that of the vertical direction, $X_3$.  The spectra are integrated over the azimuthal angle and normalised such that $\int E(k_H, k_V) dk_Hdk_V = 1$.  The black dotted lines indicate $N \sin \phi = \omega$ and $N \sin \phi = \omega / 2$, where $\phi$ is the angle of the wavevector against the vertical axis.
}
  \label{fig:result}
\end{figure}

We demonstrate the simulation results of a specific case, $Fr = 0.4$, to reveal the basic mechanisms of turbulence enhancement in the model.  The time evolutions of $\mathcal{E}_K$, $\mathcal{E}_P$, and their sum are shown in figure \ref{fig:result}a.   After a rapid decay within a couple of wave periods of both the kinetic and available potential energies due to viscosity and diffusivity, they are exponentially enhanced while periodically exchanging energy.  The step-like behavior of the total energy results from the oscillation in its production rates, $\mathcal{P}_K + \mathcal{P}_P$.  At around $t \sim 18T$, the enhancement is retarded and will gradually reach a saturation level.

Figure \ref{fig:result}b-d show snapshots of the buoyancy perturbation $b'$ on the surface of the calculation domain at $t = 15T, 18T$, and $25T$.  A notable feature is the striped pattern appearing during the stage of exponential energy growth (figure \ref{fig:result}b).  To discuss the mechanism of energy enhancement, we show in figure \ref{fig:result}e-f the energy spectra in the horizontal and vertical section at each stage.  According to the dispersion relation of internal gravity waves, the angle of the wave vector against the vertical axis $\phi$ determines the wave frequency as $\sigma = N \sin \phi$.  In figure \ref{fig:result}e, we find energy concentration along the line $N \sin \phi = \omega / 2$.  This result indicates that the subharmonic component of the outer wave is selectively excited and, consequently, the striped pattern that is perpendicular to these wave vectors is made visible.  Going back to figure \ref{fig:result}a, one may notice the exchange of $\mathcal{E}_K$ and $\mathcal{E}_P$ occurring with the same frequency as that of the outer wave, which indicates the oscillatory motion of disturbances with half the frequency of the outer wave.  All of these features are evidence that disturbance waves are excited by PSI.

\begin{figure}
  \centerline{\includegraphics[bb=0 0 454 193, width=0.9\textwidth]{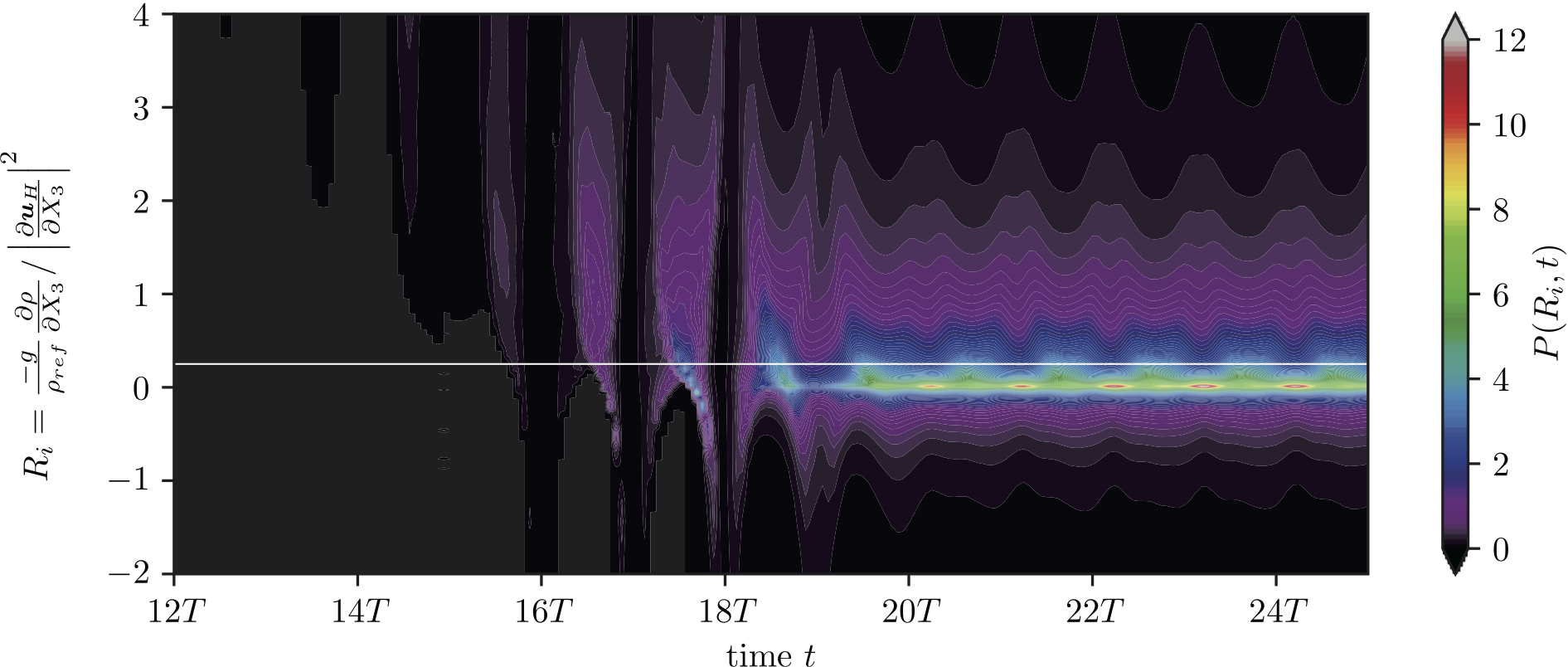}}
\caption{The time series of the probability density function of the gradient Richardson number, $R_i$, for the $Fr=0.4$ experiment.  The probability density function, $P(R_i, t)$, is normalised such that $\int P(R_i, t) d R_i = 1$.  The white horizontal line indicates $R_i = 0.25$.}
  \label{fig:Ri}
\end{figure}

To investigate the stability of PSI-induced disturbance waves, we analyse the gradient Richardson number,
\begin{equation}
R_i = \frac{ - \displaystyle \frac{g}{\rho_{ref}} \frac{\partial \rho}{\partial X_3} }{ \left| \displaystyle \frac{\partial \bm{u}_H}{\partial X_3} \right|^2 } ,
\end{equation}
where $\bm{u}_H = (u_1 \cos\theta + u_3 \sin\theta, u_2)$ is the horizontal velocity vector.  In general, the sufficient condition for shear instability is that $0 < R_i < 0.25$ is satisfied somewhere in the domain while static instability occurs when $R_i < 0$.  Figure \ref{fig:Ri} shows the time series of the probability density function (PDF) of $R_i$.  Around $t = 15.5T$, the values of $R_i$ reach lower than 0.25 and even negative values.  A short time later in $17.5T < t < 18T$, higher concentration of PDF below 0.25 becomes seen.  At this stage, secondary instabilities manifest as undulating patterns in the buoyancy field (a result at $t=18T$ is shown in figure \ref{fig:result}c).  \cite{koudella_instability_2006} argued that, for parametrically excited internal waves, the growth rate of static instability is larger than that of shear instability.  According to their theoretical prediction, the fastest growing mode of static instability would be vortices the axes of which are aligned to the direction of the sheared flow.  However, in the present experiment, as well as such predicted patterns, undulations whose crests are perpendicular to the direction of the sheared flow are visible.  We further point out that this pattern is similar to that generated by the Kelvin-Helmholtz instability.  From these results, it is speculated that both the density overturn and the strong shear cooperatively work to break the disturbance waves.  As a result, the energy accumulated in the low-wavenumber region is abruptly transferred to high-wavenumber components (figure \ref{fig:result}f).

After a sufficiently long time (figure \ref{fig:result}d, g), a smooth and nearly isotropic energy spectrum is formed, except for the lowest-wavenumber region, where anisotropic structures are maintained such that the PSI continues to supply energy into the system.  At this stage, velocity shear is dominated by the highest-wavenumber components that are hardly affected by stratification.  That is why the probability density of $R_i$ is concentrating very close to $R_i=0$ (figure \ref{fig:Ri}).  For a better visualisation, please also see the supplementary movie that shows buoyancy perturbation on the domain surface across the phases of the onset of PSI, secondary instabilities, and the fully developed turbulence.

Even in other experiments with various $Fr$, the process is essentially the same; first, specific low-wavenumber components are selectively amplified and, next, secondary instability redistributes energy across spectral space to the smallest scales.

\section{Discussion}
\begin{figure}
  \centerline{\includegraphics[bb=0 0 576 216, width=\textwidth]{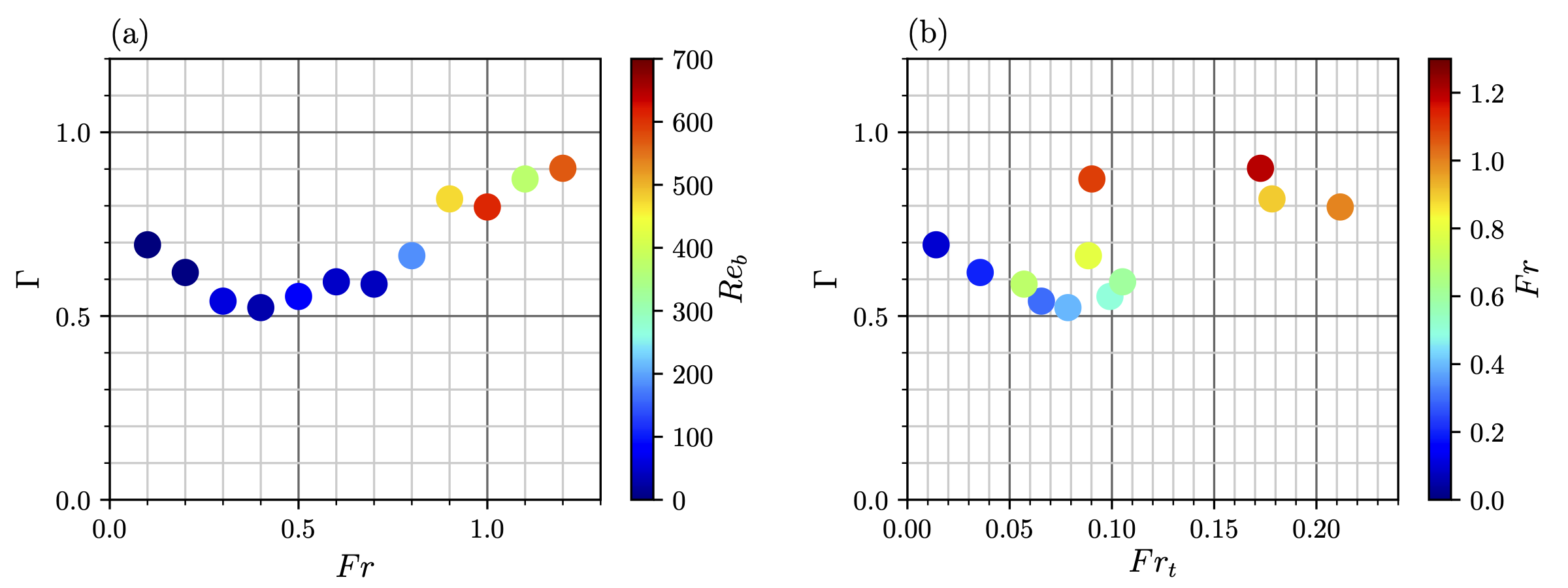}}
\caption{(a) Mixing coefficient $\Gamma \equiv \epsilon_P / \epsilon$ as a function of (external) Froude number $Fr \equiv S_0/N$.  The color represents turbulent Froude number $Fr_t \equiv \epsilon / (N \mathcal{E}_K)$.  (b) Mixing coefficient $\Gamma$ as a function of turbulent Froude number $Fr_t$.  The color represents buoyancy Reynolds number $Re_b \equiv \epsilon / (\nu N^2)$.  Here, the energy dissipation rates $\epsilon$ and $\epsilon_P$ and the kinetic energy density $\mathcal{E}_K$ are obtained by averaging each over $t_{\rm end} - T < t \leq t_{\rm end}$.}
  \label{fig:Gamma}
\end{figure}

The relative importance between the production rates of available potential energy and kinetic energy is quantified by evaluating $\mathcal{P}_P / (\mathcal{P}_K + \mathcal{P}_P)$.  As shown in Table 1, this ratio is always greater than 0.5.  Thus, we derive $\mathcal{P}_P > \mathcal{P}_K$; i.e., the instability mainly causes the production of available potential energy.  We next analyse the fraction of dissipation rates of available potential energy in the total energy loss, or the so-called mixing efficiency, $\epsilon_P / (\epsilon + \epsilon_P)$.  This value is, on the other hand, always less than 0.5 so that $\epsilon > \epsilon_P$ holds; i.e., the kinetic energy dissipation rate overwhelms the available potential energy dissipation rate.  These contrasting results indicate that part of available potential energy injected from the outer wave is converted into the kinetic energy and dissipated in heat.  Figure \ref{fig:Gamma}a shows the mixing coefficient $\Gamma = \epsilon_P / \epsilon$ as a function of $Fr = S_0/N$.  Compared to $\Gamma = 0.2$ that has been traditionally assumed for the ocean \citep{gregg_mixing_2018}, the present results that always hold $\Gamma > 0.5$ are substantially larger.  Except for $Fr \leq 0.2$ cases, $\Gamma$ tends to increase in accordance with $Fr$ and becomes close to 1 in $Fr > 1.0$ cases. 
It should be kept in mind that, when $Fr$ is small, effects from viscosity and diffusivity become relatively large, possibly leading to the property of small-scale turbulence being much different from the large $Fr$ cases.  In general, effect from viscosity to stratified turbulence can be judged from buoyancy Reynolds number, $Re_b = \epsilon / (\nu N^2)$, which is equivalent to $(\eta_O/\eta_K)^{4/3}$.  As shown in Table 1, in the cases of $Fr = 0.1$ and $0.2$, $Re_b$ is of the order of unity, so that the inertial subrange where isotropic turbulence exists collapses.  This peculiarity might cause the exceptional behavior of $\Gamma$ for these cases.

Recently, by analysing various DNS experimental data, \cite{maffioli_mixing_2016} and \cite{garanaik_inference_2019}, hereafter GV19, suggested that the turbulent Froude number, defined as $Fr_t \equiv \epsilon / (N U^2)$, where $U$ is the typical velocity of turbulence, would be a good indicator of the mixing coefficient.  According to GV19, a state of stratified turbulence is classified into three categories; for $Fr_t \gg 1$, $\Gamma \propto Fr_t^{-2}$, for $Fr_t \sim O(1)$, $\Gamma \propto Fr_t^{-1}$, and for $Fr_t \ll 1$, $\Gamma \propto Fr_t^{0}$.  In addition, their scatterplot indicates that when $Fr_t$ is sufficiently small, $\Gamma$ seems to increase slightly with $Fr_t$.  In this study, by regarding $\mathcal{E}_K$ as $U^2$, we estimate the turbulent Froude number in each run.  As figure \ref{fig:Gamma}b shows, $Fr_t$ is no more than 0.21 and $\Gamma$ exhibits a slight increasing trend with $Fr_t$, which is partly consistent with the previous results.  More classically,  $Re_b$ that involves the effect from viscosity is known to be a candidate for the indicator of $\Gamma$ \citep{gregg_mixing_2018}.  Figure \ref{fig:Gamma}a also exhibits a tendency for $\Gamma$ to increase with $Re_b$, but since $Re_b$ is inevitably correlated with $Fr$ in the current settings, causality is still uncertain.  Here, we caution that, since the vertical density gradient as well as $\epsilon$ and $\mathcal{E}_K$ vary periodically in time, instantaneous values of $Re_b$ and $Fr_t$ fluctuate by several factors.  Averaging over a wave period is essential to get a robust estimate of scaling relationships.

\begin{figure}
  \centerline{\includegraphics[bb=0 0 559 208, width=\textwidth]{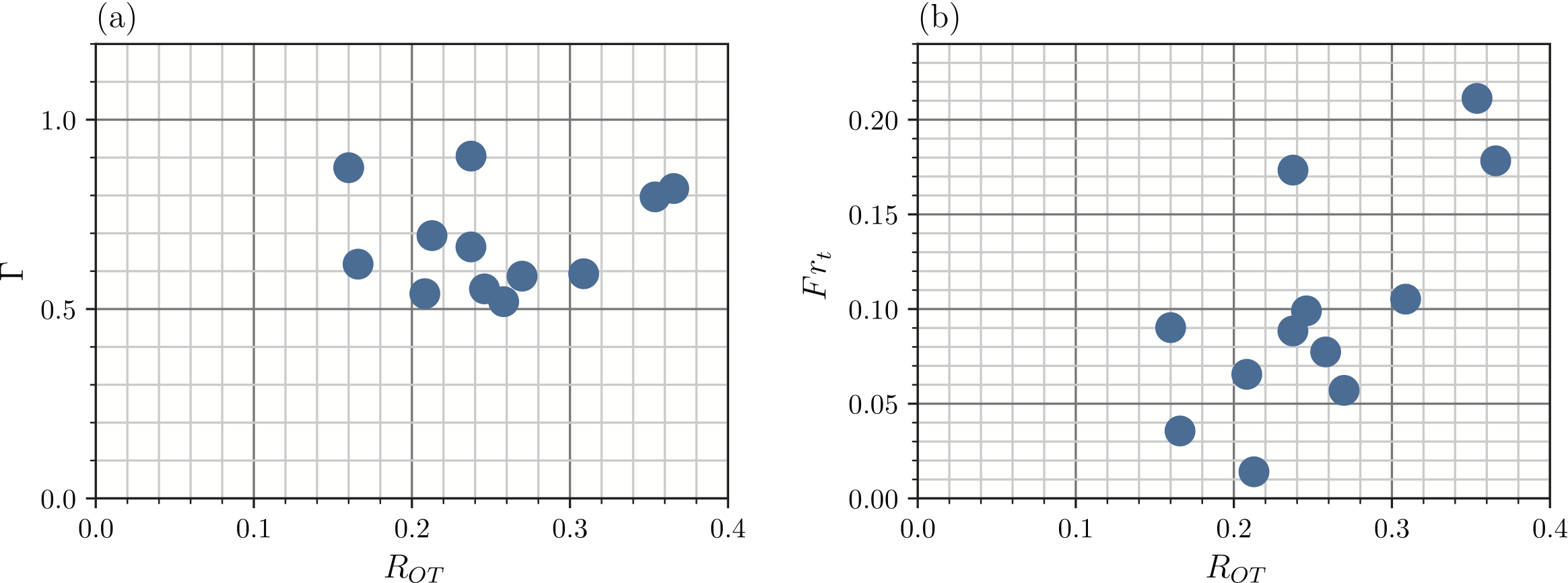}}
\caption{(a) Mixing coefficient $\Gamma \equiv \epsilon_P / \epsilon$ as a function of the ratio of the Ozmidov scale to the Thorpe scale, $R_{OT} = \eta_O / \eta_T$.  (b) Turbulent Froude number $Fr_t \equiv \epsilon / (N \mathcal{E}_K)$, as a function of $R_{OT}$.  Here, the energy dissipation rates $\epsilon$ and $\epsilon_P$ and the kinetic energy density $\mathcal{E}_K$ are obtained by averaging each over $t_{\rm end} - T < t \leq t_{\rm end}$.  To estimate the Thorpe scale, we extract 800 samples of vertical density profile data within $t_{\rm end} - T < t \leq t_{\rm end}$.}
  \label{fig:OT}
\end{figure}

\cite{ijichi_observed_2018} reported that the mixing efficiency in the deep ocean depends on the ratio of the Ozmidov scale to the Thorpe scale, $R_{OT} = \eta_O / \eta_T$.  The Thorpe scale, $\eta_T$, is the typical size of density overturns in stratified turbulence.  One can estimate $\eta_T$ by sorting instantaneous vertical density profiles and calculating the root-mean-square value of the displacements of fluid particles between sorted and unsorted profiles.  In this study, we extract 800 samples of vertical density profiles within $t_{\rm end} - T < t \leq t_{\rm end}$ from each run to calculate $\eta_T$.  From a scatterplot of $R_{OT}$ and $\Gamma$ shown in figure \ref{fig:OT}a, we cannot confirm a clear functional relationship between them.  On the other hand, as shown in figure \ref{fig:OT}b, we find a significant positive correlation between $R_{OT}$ and $Fr_t$.  In fact, if $\eta_T$ is identified with the Ellison scale, $\eta_E$, these results are again consistent with the arguments by GV19, who found the scaling relationships of $Fr_t \propto (\eta_E / \eta_O)^{-2}$ and $\Gamma \propto (\eta_E / \eta_O)^0$ for $Fr_t \ll 1$.  We note that the scaling of \cite{ijichi_observed_2018} is consistent with that of GV19 for $Fr_t \gg 1$.

Overall, our results agree with GV19 in the point of scaling relationships between $\Gamma$, $Fr_t$ and $R_{OT}$.  However, compared to the saturation level of $\Gamma \sim 0.5$ for the regime of $Fr \ll 1$ shown by GV19, those we obtain here are larger by a factor of about two.  A difference in the situations between previous studies and ours is the source of turbulence energy.  In GV19, they used three kinds of data sets of DNS experiments: freely decaying turbulence, forced turbulence, and sheared stratified turbulence.  In every case, energy is initially injected to kinetic energy and only in a second stage part of it is converted to available potential energy to mix the stratification.  In the present case, on the other hand, kinetic energy and available potential energy are simultaneously enhanced by PSI and the latter is directly used for mixing.  From the viewpoint of energetics, this process is intermediate between the conventional shear instability that is caused by the conversion from mean to turbulent kinetic energy and the convection caused by release of available potential energy.  Taking into account the fact that in free convection the mixing coefficient occasionally exceeds 1.0 \citep{davies_wykes_efficient_2014}, we can conclude that the result in our experiments, $0.5 < \Gamma < 1.0$, which connects the regimes of vertical shear-induced mixing and convectively driven mixing, is reasonable.

\section{Concluding remarks}
In this study, we have developed a new technique for direct numerical simulation (DNS) of stratified turbulence generated by instability of internal gravity waves.  A novelty of our method is to distort the model domain obliquely and periodically to simulate a turbulence enhancement due to unresolved large-scale internal waves while fully resolving the smallest-scale energy dissipation range.  We use the term ``local instability'' to represent the exponential amplification of disturbance energy in an asymptotically large-scale internal wave.  Attention is paid to the small-Froude number regime where the parametric subharmonic instability (PSI) excites a striped pattern of disturbance waves.  When the amplitude of the disturbance waves grows sufficiently large, density overturn and strong shear induce secondary instabilities that transfer energy to much smaller-scale fluctuations, resulting in an efficient turbulent mixing.

The scaling relationship between the turbulent Froude number, $Fr_t$, and the mixing coefficient, $\Gamma$, recently proposed by \cite{garanaik_inference_2019} is tested.  Our result is in part consistent with their argument; for $Fr_t  \ll 1$, $\Gamma$ is $O(1)$ and may slightly depend on $Fr_t$.  However, we have obtained a surprisingly high mixing efficiency; the value of $\Gamma$ here is larger by a factor of about two than the previous ones.  This discrepancy might be related to the difference in energy source.  We suppose that when the available potential energy of turbulence is directly supplied from large-scale fluid motion rather than converted from turbulent kinetic energy, it is more efficiently used for vertical mixing.  This thought agrees with \cite{ijichi_how_2020}, who found values as large as $\Gamma = 0.8$ near the sea floor in the Brazil Basin where hydraulic overflows are thought to cause convection, and also with the latest DNS study by \cite{howland_mixing_2020}.  In conclusion, we should not consider that the mixing efficiency is parameterised solely from a single parameter, $Fr_t$, but always pay attention to the mechanisms of turbulence generation.

Throughout this study, we have assumed that the turbulence is continuously driven by a harmonically oscillating wave shear that maintains constant amplitude.  In the real ocean, a situation where coherent internal tides dominate the internal wave spectra may satisfy this condition.  Investigating mixing caused by rather sporadic wave radiation from surface wind force requires reexamination of the model configuration.  Besides, since our model does not include the Coriolis force, the present scenario of turbulence generation and large values of $\Gamma$ may apply to limited cases when the wave frequency is much higher than the local inertial frequency, i.e., internal tides in the equatorial area.  To discuss wave-driven mixing in mid and high latitudes, we should not neglect the rotation because it will change the partitioning of kinetic and available potential energies.  Specifically, a linear theory of internal inertia-gravity waves yields the following relationship; (kinetic energy)/(available potential energy) $= (1 - \omega^2 / N^2)(\omega^2 + f^2) / (\omega^2 - f^2)$, where $\omega$ is the wave frequency and $f$ is the Coriolis parameter \citep{polzin_toward_2011}.  Consequently, as for near-inertial waves, $\omega \sim f$, a large part of energy is contained in the kinetic part so that the mixing efficiency is expected to be lowered.  In the real ocean, it has been reported that PSI plays a special role in transferring energy of internal tides into near-inertial waves \citep{hibiya_latitudinal_2004}.  Therefore, in order to get a better insight into the wave-driven ocean mixing, it is vital in future research to examine how rotation affects the energy budgets of both the internal waves and turbulence.

\section*{Acknowledgements}
The authors express their gratitude to three anonymous reviewers for their invaluable comments on the original manuscript.
This research was supported by JSPS KAKENHI Grant JP18H04918 and JP20K14556.  This work was supported by the grant ANR-17-CE30-0003 (DisET) and by a grant from the Simons Foundation (651475, TD).  Numerical calculation was conducted using the Fujitsu PRIMERGY CX600M1/CX1640M1 (Oakforest-PACS) at the Information Technology Center of the University of Tokyo.  

\section*{Declaration of interests}
The authors report no conflict of interest.

\bibliographystyle{jfm}
\bibliography{JFM2020.bib}

\end{document}